\documentclass[preprint,tightenlines,a4paper,showpacs]{revtex4}

\usepackage{amssymb}
\usepackage{latexsym}
\usepackage{amsmath}
\usepackage{graphicx}
\usepackage{bm}
\usepackage{dcolumn}

\begin{document}

\title{A New Method for Studying the Topology of the Universe\\
from the Cosmic Microwave Background}
\author{Reuven Opher}
\email{opher@astro.iag.usp.br}
\affiliation{IAG, Universidade de S\~ao Paulo, Rua do Mat\~ao 1226,\\
Cidade Universit\'aria, CEP 05508-900 S\~ao Paulo, SP, Brazil}
\date{\today}

\begin{abstract}
The standard method for observationally confirming the existence of a
predicted finite topology of the universe involves searching for
the repetition of the same finite or extended source in different
directions. However, serious problems are encountered when studying
both types of sources, finite and extended. In studying a finite
source, such as a star, galaxy, or quasar, the problem of its evolution
must be dealt with. The same source seen in different
directions is observed at different distances (i.e., ages), making its 
identification extremely difficult. Studying extended sources, such as
circles-in-the-sky (CIS) within the cosmic microwave background (CMB),
is also problematic since it requires an unrealistic zero width last 
scattering surface (LSS) to produce the predicted identical temperatures
on the circles. It is shown here that these temperatures are no longer identical
when a realistic finite width LSS is taken into account. A new method for 
studying the topology of the universe which avoids the above mentioned problems
in studying both finite and extended sources is suggested here. It consists
of searching for increased temperature fluctuations
in the regions of the intersections of the LSS with the faces of the
fundamental polyhedron of the finite topology. Appreciably greater
fluctuations are predicted in these regions. A worked-out example is given.
\end{abstract}

\maketitle

\section{INTRODUCTION}
\label{sec:intro}
Cosmic Topology aims to answer the questions as to whether the universe is
finite or infinite and the nature of its shape. In the light of existing
observational data, the three homogeneous and isotropic solutions of
Einstein's equations are the best approximations with which to describe our
universe. Each solution gives a possible local geometry of space with
constant curvature: $k=0, +1,\, \mbox{or\,}-1,$ described by the
Friedmann-Lema\^{\i}tre-Robertson-Walker (FLRW) cosmological model. Although
the FLRW solutions describe the local geometry of space, they do not fix its
topology. In general, the topology of the spatial sections of these
solutions have been taken to be trivial. For example, the case where $k=0$
is usually taken to be the infinite Euclidean space $E^{\,3}.$
\par
In non-trivial topologies, identical copies of a fundamental
polyhedron (FP) which recovers these spaces are observed. The FP is the real space in
which we live. An illusion of multiple copies of the FP and repeated images of
its sources arises from the identification of pairs of its faces~\cite{lalu}. 
Because light from a source travels in all directions, crossing the faces
of the FP, the illusion of observing the same source as well as multiple copies 
of the FP from different directions is created.
\par
Until now, the methods used to look for signs of non-trivial topologies have been
based on the presence of multiple images of the same finite or extended source. These methods
were developed by analyzing temperature distributions in the
cosmic microwave background (CMB) radiation (for extended sources)~\cite{css98,JJ98,jl} 
or the three-dimensional distributions of cosmic
sources (for finite sources)~\cite{lalu,lelalu,R96,H89}. In recent
years, several of these methods have been extended to the three geometries
with constant curvature, $k=0,+1,-1,$ and their respective non-trivial
topologies~\cite{GLLUW, LWUGL,HG98,HG99}.
\par
One of the methods suggested for trying to detect non-trivial topologies in
the CMB was the circles-in-the-sky (CIS) method. This method is based on searching
for circles in the CMB with identical temperature distributions~\cite{css98}.
In a universe with a non-trivial topology, the LSS is seen to be repeated in each
copy of the FP which is observed. If the dimensions of the FP observed are less than or on the order of
the LSS, copies of the LSS will intersect with one another. These
intersections will then produce matched circle pairs, where the temperature
distributions along the circles are predicted to be identical. The pattern of matched circles will vary
according to the topology. However, as noted above, the observed multiple copies of the FP are, in fact, 
actually an illusion due to the fact that light travels in all directions, crossing the faces of the FP.
\par
Recently, cosmic topology aroused great interest, when Luminet et al.~\cite{nat} 
suggested that the WMAP data indicate that we live in a universe with
Poincar\'e dodecahedron space. However, it was subsequently argued by
Cornish et al.~\cite{cssk03} that, when applied to the WMAP data, the CIS
analysis excludes the Poincar\'e dodecahedron space.
\par
An essential condition is imposed
on the LSS by the CIS method for studying topology, namely, that the
transition of an opaque to a transparent universe occurs instantaneously.
Therefore, the shell, from where the observed photons came, should be
infinitesimally thin~\cite{wc}.
\par
However, decoupling of the photons in the early universe was, in fact, not an
instantaneous event, as assumed in the CIS method, but took place in a
redshift interval $\Delta z_{\,\textrm{LSS}}.$ The probability that an
observed photon was last scattered in the redshift interval 
$\Delta z_{\,\textrm{LSS}}$ around an average $z_{\,\textrm{LSS}}$ can be approximated by a
Gaussian distribution with a mean value $z_{\,\textrm{LSS}}$ and a width 
$\Delta z_{\,\textrm{LSS}}.$ A reshift interval $\Delta z_{\,\textrm{LSS}} \sim
80,$ for example, corresponds to a proper length $\Delta l\sim 15\Omega_0
^{-1/2}\;\textrm{Mpc}$ and an angle $\Delta l\sim 8^{\prime}\Omega ^{1/2}h,$
where $h=H_{0}/100\,\textrm{km\,s}^{-1}\,\textrm{Mpc}^{-1},$ $H_{0}$ is the
Hubble constant, and $\Omega_0$ is the density parameter \cite{pad}. The
density parameter for the recent WMAP data is $\Omega_0=1.02\pm 0.02.$
\par
The problem with observing multiple images of a single source, such as a
star, galaxy, or quasar, is the time evolution of the source. For example,
if the dimension of the FP is on the order of the horizon, the difference in
the ages of the same source is on the order of the age of the universe.
Thus, an observed source at redshift $z\sim 5$ may have an age $\sim 10$
million years, while the same source observed at $z\sim 0$ may have an age 
$\sim 10$ billion years. Obviously, the two images will not look the
same. Consequently, observing multiple images of a single source, such as a
star, galaxy or a quasar, must take into account the extremely difficult problem
of identifying the source, which is evolving in time.
\par
Instead of trying to observe the same source from different directions,
where time evolution must be taken into account, or searching for identical
CBR temperatures on separate circles, where a zero, nonrealistic width
LSS $(\Delta z_\textrm{LSS}=0)$ is assumed, as in the CIS method, we suggest observing
the regions where the LSS intersects the face of the FP, in which we predict an
appreciable increase in the temperature fluctuations. The new method suggested 
here avoids the problems mentioned above. It is based on a realistic finite
width of the LSS and does not involve the problem of time evolution. In
Section~\ref{sec:top}, we describe the method and give a worked-out
example. Conclusions and discussion are given in Section~\ref{sec:cd}.

\section{TOPOLOGY OF THE UNIVERSE WITH A FINITE WIDTH LSS}
\label{sec:top} 
Let us observe the CMB in a direction where the LSS
intersects the face of the FP at the mean redshift $z_{\,\textrm{LSS}},$
which has a width $\Delta z_{\,\textrm{LSS}}.$ We define the plasma of the LSS
between us and $z_{\,\textrm{LSS}}$ (i.e., between the redshifts 
$z_\textrm{MIN}=z_{\,\textrm{LSS}}-(\Delta z_{\,\textrm{LSS}})/2$ and $z_{\,\textrm{LSS}})$ as 
$P_1.$ The temperature $T_1$ of $P_1$ is different from $T_2,$ the
temperature of the plasma $P_2$ between $z_{\,\textrm{LSS}}$ and $z_{\,\textrm{MAX}}$
(i.e., between the redshifts 
$z_{\,\textrm{LSS}}$ and $z_{\,\textrm{MAX}}=z_{\,\textrm{LSS}} + (\Delta z_{\,\textrm{LSS}})/2).$ 
This difference in temperature is due to density fluctuations in the primordial plasma.
Since $z_{\,\textrm{LSS}}$ is at the face of the FP, both $P_1$ and $P_2$ are
again observed in the opposite direction in the FP, where $P_2$ is now
observed before $P_1.$ Clearly, the temperature fluctuations created by $P_1$ and 
$P_2$ are not independent since the same plasmas are observed in
both directions.
\par
We now observe the CMB in two different directions in the case where the LSS is within the FP and
does not intersect its faces. Here, the temperature fluctuations
are independent since the plasma observed in one direction is not repeated
in the other direction.
\par
The average temperature fluctuations observed in case I, where the LSS
intersects the faces of the FP, will be compared with those in case II,
where the faces of the FP are not intersected by the LSS. From the following
calculations, it is seen that the temperature fluctuations in the first case
are appreciably greater than those in the second.
\par
In case I, an observer looking at one side of the FP in the direction of
face A measures the intensity of the radiation from $P_1$ and $P_2$ in terms
of an effective temperature $T_{12},$ which is a weighted mean of the
temperatures of $P_1(T_1)$ and $P_2(T_2):$ 
\begin{equation}
T_{12}=g_{\,\textrm{NEAR}}\,T_1+g_{\,\textrm{FAR}}\,T_2,  
\label{eq:one}
\end{equation}
where $g_{\,\textrm{NEAR}}$ is the probability that an observed photon came
from the near plasma, from $z_{\,\textrm{MIN}}$ to $z_{\,\textrm{LSS}},$ and 
$g_{\,\textrm{FAR}}$ is the probability that it came from the far plasma, from 
$z_{\,\textrm{LSS}}$ to $z_{\,\textrm{MAX}}.$ Since photons in the interval 
$z_{\,\textrm{MIN}}$ to $z_{\,\textrm{LSS}}$ need to pass through only $\sim 1/4$ of
the thickness of the LSS, $g_{\,\textrm{NEAR}}\sim 0.75.$ On the other hand, 
$g_{\,\textrm{FAR}}$ is relatively small, $\sim 0.25,$ since photons in the
redshift interval $z_{\,\textrm{LSS}}$ to $z_{\,\textrm{MAX}}$ need to pass
through $\sim 3/4$ of the LSS. The illusion of multiple
copies of sources arises from the identification of pairs of 
faces of the FP (Section~\ref{sec:intro}). We not only observe $P_1$ 
in the redshift interval $z_{\,\textrm{MIN}}$
to $z_{\,\textrm{LSS}}$ in the direction of face A, but
also in the redshift interval $z_{\,\textrm{LSS}}$ to $z_{\,\textrm{MAX}}$ in
the direction of face B. Likewise, we observe $P_2$ in the redshift interval 
$z_{\,\textrm{LSS}}$ to $z_{\,\textrm{MAX}}$ in the direction of face A as well
as in the redshift $z_{\,\textrm{MIN}}$ to $z_{\,\textrm{LSS}}$ in the direction
of face B. Thus, looking at the opposite side of FP in the direction of face
B, the radiation from $P_1$ and $P_2$ in terms of an effective temperature 
$T_{21},$ is now 
\begin{equation}
T_{21}=g_{\,\textrm{NEAR}}\,T_2+g_{\,\textrm{FAR}}\,T_1.  
\label{eq:two}
\end{equation}
The sum of the temperatures ${T_S}_{\textrm{I}}$ from $T_{12}$ and $T_{21}$ in case I
is 
\begin{align}  
\label{eq:three}
{T_S}_{\textrm{I}}& =g_{\,\textrm{NEAR}}T_1+g_{\,\textrm{FAR}}T_2+g_{\,\textrm{NEAR}}T_2+
g_{\,\textrm{FAR}}T_1 \\
&= [g_{\,\textrm{NEAR}}+g_{\,\textrm{FAR}}][T_1+T_2].  \notag
\end{align}
Both $T_{12}$ and $T_{21}$involve two plasma regions. Thus, the sum ${T_S}_{\textrm{I}}$ involves
four terms. However, only two regions are independent since the other two are
repetitions of the first.
\par
When the LSS is within in the FP and doesn't intersect its faces, the non-trivial
topology is not evident. Thus the temperature sum in two different directions involves
four different plasma regions of the LSS, instead of only two. The sum of the temperatures 
${T_S}_{\textrm{II}}$ from the two directions in case II is 
\begin{equation}
{T_S}_{\textrm{II}}=g_{\,\textrm{NEAR}}T_1+g_{\,\textrm{FAR}}T_2+g_{\,\textrm{NEAR}}T_3+
g_{\,\textrm{FAR}}T_4  
\label{eq:four}
\end{equation}
Because this temperature sum involves four different regions and not just
two, as in ${T_S}_{\textrm{I}},$ this temperature sum has smaller fluctuations, as we
shall now show.
\par
Let us assume that each different plasma region, from $z_{\,\textrm{MIN}}$ to 
$z_{\,\textrm{LSS}}$ and from $z_{\,\textrm{LSS}}$ to $z_{\,\textrm{MAX}},$ has a
random probability of having a temperature $T_0+\delta T$ or $T_0-\delta T,$
where $T_0$ is the mean CMB temperature.
From measurements of the regions near the intersection of the LSS with the faces
A and B of the FP in case I, we obtain a probability distribution function
of obtaining a temperature ${T_S}_{\textrm{I}}=2T_0 + \Delta T_1,\; 2T_0+\Delta T_2,\;
2T_0+\Delta T_3,\;\textrm{etc}.$ The values 
$\Delta T_1,\;\Delta T_2,\;\Delta T_3,\;\textrm{etc.}$ describe a probability
distribution function whose width at half height is  
\begin{equation}
\overline{\Delta T_{\textrm{I}}}=\left [g_{\,\textrm{NEAR}}+g_{\,\textrm{FAR}}\right] 2\delta T\simeq 2\delta T.  
\label{eq:five}
\end{equation}
In the same way, the value of the width at half height of the
probability distribution function in case II is 
\begin{align}  
\label{eq:six}
\overline{\Delta T_{\textrm{II}}} &= 0.84\left[g_{\,\textrm{NEAR}}+g_{\,\textrm{FAR}}\right]2\delta 
T_{\textrm{II}} \\
&\simeq 0.84\times (2\delta T_{\textrm{II}}). \notag
\end{align}
Thus, $\overline{\Delta T_{\textrm{I}}}$ has a value which is 16\% greater than $\overline{\Delta T_{\textrm{II}}}.$

\section{CONCLUSIONS AND DISCUSSION}
\label{sec:cd}
We suggest here the comparison of temperature fluctuations near the intersections of
the LSS and a pair of faces of the FP with those
taken where the LSS does not intersect the faces of the FP. The
temperature fluctuations are predicted to be appreciably different. In the
worked-out example given, a 16\% difference was found.
\par
It is interesting to note from Eqs. (\ref{eq:one}) and (\ref{eq:two}) that,
in general $T_{12}\ne T_{21}.$ This indicates a weakness of the CIS method.
\par
We strongly suggest that the new method for observing the
topology of the universe, discussed here, be explored with present and future CMB data. A
telescope with a resolution on the order of the thickness of the LSS should
be used. The recent WMAP data indicates that 
$\Delta z_{\,\textrm{LSS}}\simeq 195$ and $\Omega \simeq 102$~\cite{dns}. This corresponds
to a proper thickness of the LSS $\sim 37\,\textrm{Mpc}$ and an angle $\sim
20\,\textrm{arcmin}.$ The resolution of the WMAP data is 
$\sim 42\,\textrm{arcmin}.$

\begin{acknowledgments}
R. O. would like to thank the Brazilian agencies CNPq (300414/82-0) and
FAPESP (00/06770-2) for partial support.
\end{acknowledgments}

\end{document}